\documentclass[aps,prl,reprint,showpacs,superscriptaddress]{revtex4-2}

\usepackage{graphicx}
\usepackage{bm}
\usepackage[utf8]{inputenc}
\usepackage{amsmath}
\usepackage{float}
\usepackage{amsmath} 
\usepackage{booktabs}
\usepackage[x11names,table,xcdraw]{xcolor}
\usepackage{comment}

\usepackage{soul}

\begin{document}
\preprint{APS/123-QED}
\title
{Control of domain wall and pinning disorder interaction by light He$^+$ ion irradiation in Pt/Co/AlOx ultrathin films}

\author{Cristina Balan}
\affiliation{Univ. Grenoble Alpes, CNRS, Institut N\'eel, 38042 Grenoble, France}
\author{Johannes W. van der Jagt}
\affiliation{Spin-Ion Technologies, 10 Boulevard Thomas Gobert, 91120 Palaiseau, France }
\affiliation{Centre de Nanosciences et de Nanotechnologies, CNRS, Université Paris-Saclay, 10 boulevard Thomas Gobert, 91120 Palaiseau, France 
}
\author{Jose Pe$\tilde{\mathrm{n}}$a Garcia}
\affiliation{Univ. Grenoble Alpes, CNRS, Institut N\'eel, 38042 Grenoble, France}
\author{Jan Vogel}
\affiliation{Univ. Grenoble Alpes, CNRS, Institut N\'eel, 38042 Grenoble, France}
\author{Laurent Ranno}
\affiliation{Univ. Grenoble Alpes, CNRS, Institut N\'eel, 38042 Grenoble, France}
\author{Marlio Bonfim}
\affiliation{Dep. de Engenharia Elétrica,
	Universidade Federal do Parana, Curitiba, Brasil}
\author{Dafiné Ravelosona}
\affiliation{Spin-Ion Technologies, 10 Boulevard Thomas Gobert, 91120 Palaiseau, France }
\affiliation{Centre de Nanosciences et de Nanotechnologies, CNRS, Université Paris-Saclay, 10 boulevard Thomas Gobert, 91120 Palaiseau, France 
}
\author{Stefania Pizzini}
\affiliation{Univ. Grenoble Alpes, CNRS, Institut N\'eel, 38042 Grenoble, France}

\author{Vincent Jeudy}
\affiliation{Laboratoire de Physique des Solides, Universit\'{e} Paris-Saclay, CNRS, 91405 Orsay, France}
\email{vincent.jeudy@universite-paris-saclay.fr, stefania.pizzini @neel.cnrs.fr}
\begin{abstract}
We have studied the effect of He$^+$ irradiation on the dynamics of chiral domain walls in Pt/Co/AlOx trilayers in the creep regime. The irradiation leads to a strong decrease of the depinning field and a non-monotonous change of the effective pinning barriers. The variations of domain wall dynamics  result essentially from the strong decrease of the effective anisotropy constant, which increases the domain wall width. The latter is found to present a perfect scaling with the length-scale of the interaction between domain wall and disorder, $\xi$. On the other hand, the strength of the domain wall-disorder interaction, $f_{pin}$, is weakly impacted by the irradiation, suggesting that the length-scales of the disorder fluctuation remain smaller than the domain wall width.
\end{abstract}

\maketitle


One of the most critical technological issues that hinders the application of magnetic textures such as domain walls (DWs) and skyrmions to performing spintronic devices, is their interaction with defects. Material inhomogeneities act as pinning sites for magnetic textures, limiting their velocities for small driving torques (i.e. for fields or currents below the depinning threshold, in the so-called creep regime) and preventing reproducible displacement events. To control the motion of magnetic textures, a better understanding of their interaction with the pinning disorder would be particularly welcome. 

In this frame, ion irradiation is an interesting tool since it allows both tuning the magnetic properties of ultrathin films~\cite{Chappert1998,Devolder2000,devolder_epjb_2001,Fassbender2004,Balk2017,Sud2021,Juge2021,deJong2022} and modifying DW dynamics~\cite{herrera-diez_apl_2015,vanderJagt2022}. Early studies on Pt/Co/Pt multilayers showed that He$^+$ ions with energy in the 30 keV range provoke  short-range (0.2-0.5 nm) atomic displacements through low energy collisions~\cite{Chappert1998,Ferre1999,Devolder2000}. The resulting Co/Pt intermixing gradually evolves with increasing fluences, therefore tuning the interfacial perpendicular magnetic anisotropy (PMA) \cite{Weller1995}.  Although ion irradiation has been observed to change the DW dynamics of several ultrathin films in the creep regime ~\cite{herrera-diez_apl_2015,vanderJagt2022}, the microscopic origin of these effects remains an open issue, as ion irradiation may modify both the pinning disorder and the DW magnetic texture (via the film PMA) and thus the DW-disorder interaction.  

Recent developments in the understanding of pinning dependent dynamics of DWs driven by magnetic field~\cite{jeudy_prl_2016,diaz_PRB_2017_depinning} and electrical current (via the spin transfer torque)~\cite{diaz_pardo_prb_2019} provide quantitative assessments on DW-disorder interactions. The pinning dependent dynamics of DWs results from the interplay between DW elasticity, weak pinning, thermal activation and a driving force~\cite{lemersle_prl_1998}. The thermally activated creep regime~\cite{jeudy_prl_2016} and depinning~\cite{diaz_PRB_2017_depinning} regime observed below and just above the depinning threshold present well studied universal behaviors, which are in agreement with the predictions for the quenched Edwards-Wilkinson universality class~\cite{albornoz_prb_2021}.
Combining a self-consistent description of the creep and depinning regimes and a scaling model of DW depinning, the analysis of domain wall dynamics allows extracting the
parameters characterizing the interaction between domain walls and weak pinning disorder ~\cite{gehanne_prr_2020}. 

In this work, we analyse the evolution with He$^+$ fluence of micromagnetic parameters and DW dynamics in a series of Pt/Co/AlOx ultrathin films presenting the same initial disorder. We evidence a perfect scaling between DW width parameter and DW-disorder interaction length scale, directly reflecting the strong decrease of the PMA.  On the other hand, the strong correlation between pinning strength and DW energy is compatible with a negligible effect of He$^+$ irradiation on the pinning disorder.


Ta(4)/Pt(4)/Co(1.1)/Al(2) magnetic stacks (thicknesses in nm) were deposited by magnetron sputtering on Si/SiO$_2$ wafers; the Al layer was consequently oxidized with an oxygen plasma. The film was diced into small samples: one of them was kept in the pristine state, while the others were irradiated at room temperature with increasing He$^+$ fluence, ranging from 4$\times$10$^{14}$ to 1.5$\times$10$^{15}$ He$^+$/cm$^{2}$. The measured magnetic parameters  are presented in Table I. The spontaneous magnetization $M_s$ and the anisotropy field $\mu_0 H_k$  were measured by superconducting quantum interference vibrating sample magnetometry.  All the samples present an out-of-plane easy magnetization axis. The in-plane saturation field strongly decreases as the He$^+$ fluence increases, while the spontaneous magnetization is practically unchanged.  


The field-driven domain wall dynamics was measured using polar magneto-optical Kerr microscopy.  For the lowest DW velocities, out-of-plane magnetic field pulses were applied using an electromagnet (maximum pulse amplitude $\mu_0 H_z$=25 mT, minimum duration 20 ms). For the fastest velocities, pulses of maximum amplitude $\mu_0 H_z$=200 mT and minimum duration 30 ns were delivered by a 200 $\mu$m-diameter microcoil associated to a fast pulse current generator \cite{Pham2016}. 
The film magnetization was first saturated in the out-of-plane direction. An opposite magnetic field pulse was then applied to nucleate a reverse domain.  The velocity of DWs was deduced from their displacement observed after the magnetic field pulse, and corresponds to ratio between the displacement and the pulse duration. The presence of left-hander homochiral Néel walls \cite{Thiaville2012} associated to the presence of Dzyaloshinskii-Moriya interaction (DMI) \cite{Dzyaloshinskii1957,Moriya1960} was confirmed by the non-isotropic displacement of the DWs in the presence of a static in-plane magnetic field $\mu_0 H_x$ (not shown).
The strength of the DMI interaction was obtained from the DW saturation velocity at large $B_z$ fields v$_{sat}=\gamma \pi D/(2 M_s)$ \cite{Pham2016,Krizakova2019}.   

\begin{table*}
	\caption[]{\textbf{Micromagnetic parameters measured for the Pt/Co/AlOx film in the pristine state and after irradiation with He$^{+}$ ions.}	
		For each sample, the table indicates the He$^{+}$ irradiation fluence, the spontaneous magnetization $M_s$, the anisotropy field $\mu_0 H_{K}$, the anisotropy constant $K_{eff}=\mu_0 H_{K} M_s/2$, the DW width parameter  $\Delta=\sqrt{A/K_{eff}}$, the DW saturation velocity $v_{sat}$ obtained for high magnetic fields, the interfacial DMI constant $D$  obtained using  $D= 2 M_s v_{sat}/(\gamma \pi)$ and the DW energy $\sigma=4\sqrt{AK_{eff}}-\pi D$ using $A$=16 mJ/m.}
	
	\begin{tabular}{l c c c c c c c c }\hline \\
		Sample & He$^{+}$fluence & $M_{s}$ &$\mu_{0}H_{K}$ & $K_{eff}$ & v$_{sat}$   & $\Delta$ &$D$ &$\sigma$  \\
		&   [ions/cm$^{2}$] & [MA/m ]  &    [T]  & [m/s]         &     [10$^{5}$J/m$^3$]  &  [nm] & [mJ/m$^2$] & [mJ/m$^2$] 
		  \\ \hline		
		pristine & 0  & 1.13$\pm$0.03 & 785$\pm$40 &   4.44$\pm$0.25 & 230$\pm$20 & 6.0$\pm$0.2 & 0.94$\pm$0.10 & 7.7$\pm$0.9       \\
		F1  & 4.0x10$^{14}$ & 1.13$\pm$0.03 & 727$\pm$40   & 4.11$\pm$0.24   &260 & 6.3$\pm$0.2 & 1.06$\pm$0.10 &  7.1$\pm$0.8  \\
		F2  & 1.0x10$^{15}$ & 1.16$\pm$0.03 & 446$\pm$25 & 2.59$\pm$0.15 & 260  & 7.9$\pm$0.2& 1.09$\pm$0.10&  4.9$\pm$0.6   \\
		F3  & 1.5x10$^{15}$ & 1.17$\pm$0.03 & 305$\pm$16 &  1.78$\pm$0.10 & 260 & 9.5$\pm$0.2& 1.10$\pm$0.10&  3.3$\pm$0.4  \\		
		\hline		
	\end{tabular}
\end{table*}

%
\begin{figure}[b]
	\centering
	\includegraphics[scale=0.75]{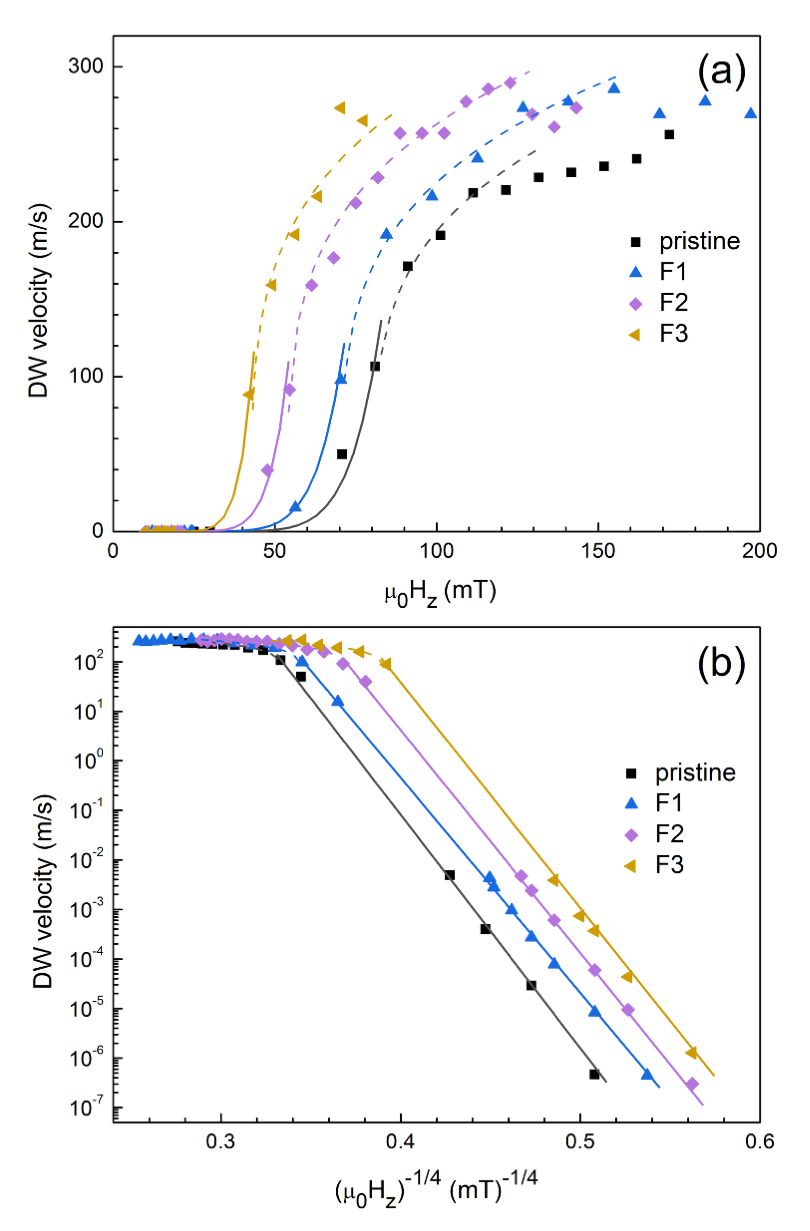}
	\caption{\textbf{Effect of He$^+$ ion irradiation on domain wall dynamics}:
		(a) domain wall velocity versus out-of plane field pulse amplitude $\mu_0 H_z$ for different irradiation fluences; (b) the same curves plotted in semi-log scale as a function of $(\mu_0 H_z)^{-1/4}$ showing the linear trend expected for the creep regime. The solid and dashed lines correspond to fits of Eqs. \ref{eq: 1} for the creep and depinning regime, respectively. } 
	\label{fig:DWdynamics}
\end{figure}

The domain wall velocities driven by  out-of-plane magnetic fields up to 200 mT are reported in Fig. \ref{fig:DWdynamics} for the samples in the pristine state and after irradiation.  As it can be observed, the strongest trend is a shift of the curves towards low field values with increasing irradiation fluence. More quantitative insights on the DW dynamics can be deduced from the self-consistent description of the creep and depinning regimes proposed in Refs. ~\cite{jeudy_prl_2016,diaz_PRB_2017_depinning,jeudy_PRB_2018_DW_pinning}: 
\begin{equation}
	v(H_z)=\left \{
	\begin{array}{lr}
		v(H_d)\exp [-\frac{T_d}{T}((\frac{H_d}{H_z})^{\mu}-1)]  & creep:H_z<H_d\\
		\frac{v(H_d)}{x_0}(\frac{T_d}{T})^\psi(\frac{H_z-H_d}{H_d})^\beta & depinning:H_z \gtrsim H_d, \\
	\end{array}
	\right.
	\label{eq: 1}
\end{equation}
where $\mu=1/4$, $\beta=0.25$, and $\psi=0.15$ are universal critical exponents and $x_0=0.65$ a universal constant~\cite{bustingorry_PRE_12_thermal_rounding, diaz_PRB_2017_depinning}. In Eqs.\ref{eq: 1}, the three adjustable parameters depend on the film magnetic and pinning properties: the depinning temperature $T_d$ characterizing the height of the effective pinning barrier, the depinning field $H_d$ and the velocity $v(H_d)$, corresponding to the coordinates of the crossover between creep and depinning. The good agreement between the experimental curves and the fit using Eqs.~\ref{eq: 1}  (Fig.~\ref{fig:DWdynamics}(a-b))
evidences the crossover between the creep and depinning regimes.
\begin{figure}[t]
	\centering
	\includegraphics[scale=0.75]{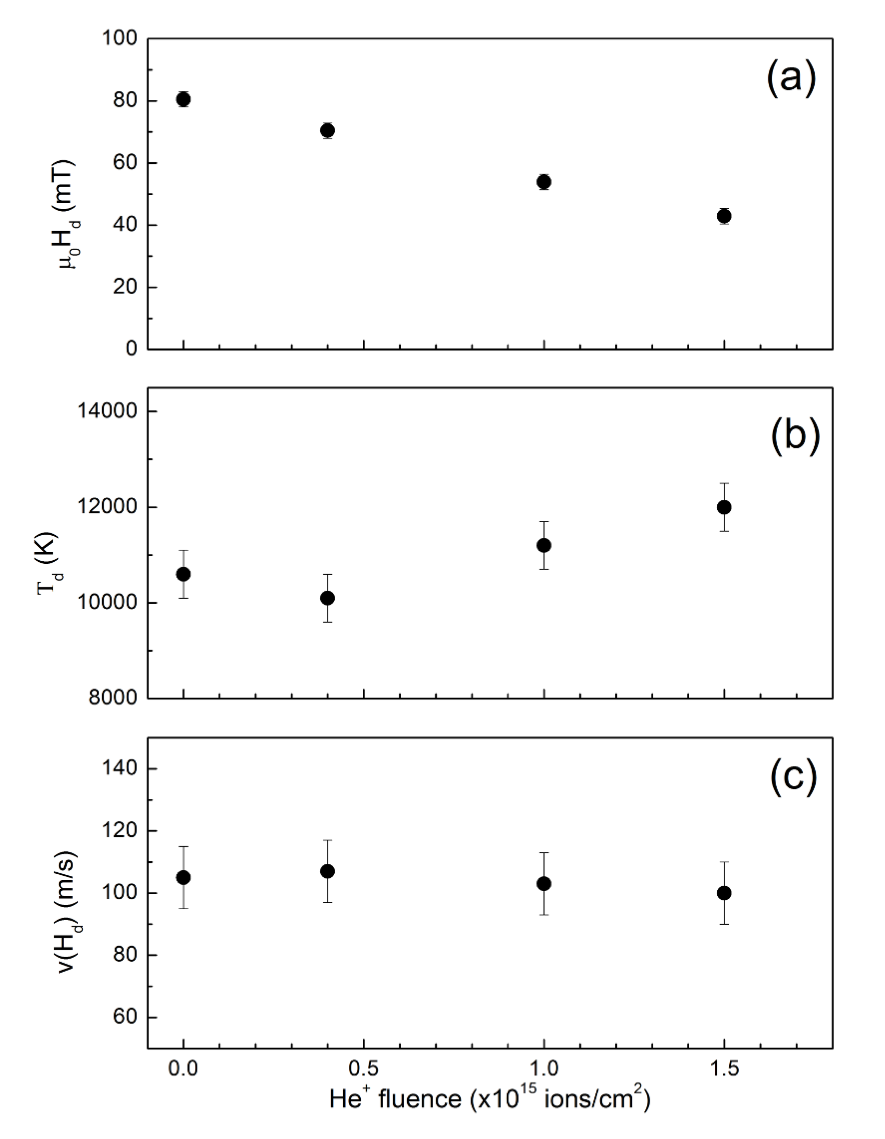}
	\caption{\textbf{Effect of He$^+$ ion irradiation on the depinning parameters}:
		(a) depinning field $\mu_0H_d$,  (b) depinning temperature $T_d$ and (c) depinning velocity $v(H_d)$. }
	\label{fig:2}
\end{figure}
The evolution of the depinning parameters with the irradiation fluence is reported in Fig.~\ref{fig:2}. The irradiation has no effect on the depinning velocity $v(H_d)$, which remains rather constant. $T_d$ varies slightly with a non-monotonous trend. In contrast, $H_d$ decreases by a factor $\approx 2$ with the irradiation fluence. (from around 80 mT for the pristine sample, down to 40 mT for 1.5$\times$10$^{15}$ He$^+$/cm$^{2}$)  As a consequence, in the irradiated samples the DWs can reach the largest velocities for lower applied magnetic field.
Note that our results are not compatible with the assumption $\sigma \sim T_dH_d^{1/4}$ proposed by Je et al. \cite{Je2013} since the slope of the creep law in Fig.~\ref{fig:DWdynamics}(b) ($\propto T_dH_d^{1/4}$) remains rather constant while the DW energy $\sigma $ (see Table 1) varies by more than a factor 2.

In order to discuss the evolution of the interaction between DWs and disorder as a function of  irradiation fluence, we use the scaling model developed in Ref.~\cite{gehanne_prr_2020}. This model allows us to link the characteristic length scale  $\xi$ and the force $f_{pin}$ of the interaction between DW and disorder to the measured depinning field $H_d$, the temperature $T_d$ and the micromagnetic parameters, i.e. the DW energy $\sigma=4\sqrt{AK_{eff}}-\pi D$ 
\cite{Thiaville2012} and the spontaneous magnetization $M_s$:
\begin{eqnarray}
	\label{eq: 3_xi}
	\xi \sim \left[ (k_BT_d)^2/(2\mu_0 H_dM_s\sigma t^2)\right]^{1/3},\\
	\label{eq: 2_fpin}
	f_{pin} \sim \frac{b}{\xi}\sqrt{2\mu_0 H_dM_stk_BT_d},
\end{eqnarray}
In Eqs.~\ref{eq: 3_xi} and \ref{eq: 2_fpin}, $A$ is the exchange stiffness, $K_{eff}$  is the effective anisotropy constant, $D$ is the DMI constant,  $k_B$ is the Boltzmann constant, $t$ is the magnetic film thickness, and $b$ the characteristic distance between  pinning sites.
%
\begin{figure}[h]
	\centering
	\includegraphics[scale=0.7]{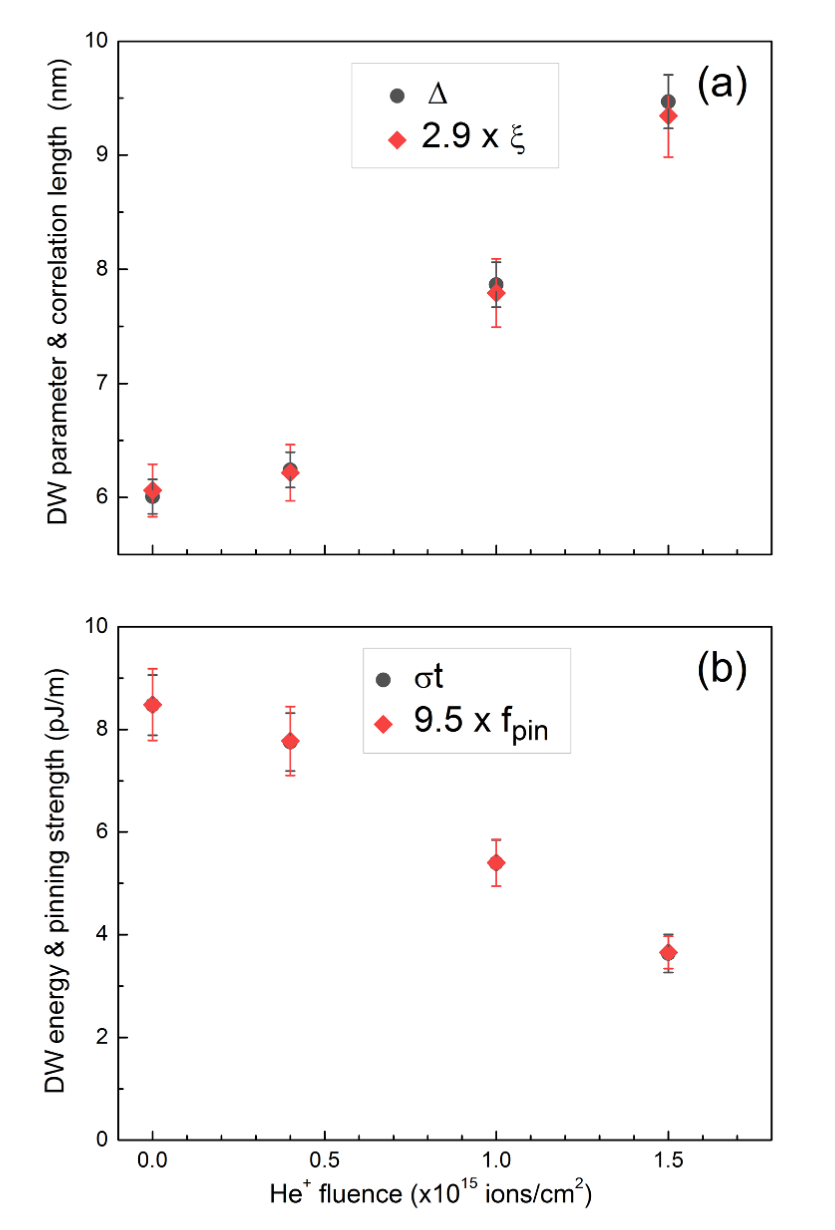}
	\caption{\textbf{Effect of He$^+$ ion irradiation on DW-disorder characteristic length scale and force}: comparison between
		(a) characteristic length scale $\xi$ and DW parameter $\Delta$, (b) pinning force $f_{pin}$ and DW energy per unit length $\sigma$.}
	\label{fig:3}
\end{figure}
Fig.~\ref{fig:3}(a) compares the evolution  of  $\xi$ and of the DW parameter $\Delta$ as a function of the irradiation fluence. As it can be observed, rescaling $\xi$ with a single adjustable parameter ($\Delta \approx 2.9 \xi$) leads to a superposition with the values of  $\Delta$. 
This almost perfect scaling strongly suggests that the characteristic distance between pinning sites $b$ is smaller than $\Delta$,  so that the latter fixes the length-scale of DW-disorder interaction~\cite{nattermann_1990_creep_law,gehanne_prr_2020}. As a consequence, the predicted short-range atomic displacements produced by He$^+$ ions irradiation~\cite{Devolder2000,devolder_epjb_2001} are shown to have no impact on the characteristic length of DW-pinning interaction $\xi$. The observed variation of $\xi$ only reflects the strong decrease of effective anisotropy constant with increasing irradiation fluence. 
%


Fig.~\ref{fig:3}(b) compares the evolution of the pinning force $f_{pin}$ and the DW energy per unit length $\sigma t$. In this case, we assumed a common distance between pinning sites $b$= 1nm, and also used a single scaling parameter ($=9.5$) to superimpose the data points corresponding to the pristine film. $f_{pin}$ and $\sigma t$ are observed to follow a similar decreasing trend  with the irradiation fluence. Note that a similar phenomenon was already observed for Pt/Co/Pt, Pt/Co/Au, and Au/Co/Pt trilayers in Ref.\cite{gehanne_prr_2020}. In that work, the samples had a fixed disorder and the DW width and energy were controlled by an in-plane magnetic field \cite{gehanne_prr_2020}. In the present case, as the samples share initially the same pinning disorder, the observed close trend of $f_{pin}$ and $\sigma t$ suggests a weak change of the disorder by irradiation. Assuming a perfect scaling between $f_{pin}$ and $\sigma t$ (i.e. adjusting the value of $b$ to perfectly superimpose the data points in Fig.~\ref{fig:3}(b)) leads a rough estimate of the maximum variation of $b$  $\approx -15\%$. 
These insights suggest that the He$^+$ irradiation has little impact on the pinning disorder and that the variation of the strength DW-disorder interaction $f_{pin}$ essentially reflects the decrease of the effective anisotropy constant with irradiation fluence. 

In conclusion, light He$^+$ irradiation in Pt/Co/AlOx ultrathin films causes to a strong reduction of the depinning field, leading to an increase the DW mobility at low fields. 
Through a self-consistent description of the creep and pinning dynamics completed by a scaling model of DW depinning (relating DW pinning properties to depinning and micromagnetic parameters), we reveal an excellent scaling between the variations of the DW-disorder interaction length scale $\xi$ and the DW width parameter.
This scaling strongly suggests that the modifications DW pinning are essentially dominated by the variations of DW magnetic texture (via the variation of the film anisotropy), while short range atomic displacement produced by irradiation have weak impact on the pinning disorder.

\begin{acknowledgments}
We acknowledge the support of the Agence Nationale de la Recherche (projects ANR-17-CE24-0025 (TOPSKY) and of the DARPA TEE program through Grant No. MIPR HR0011831554. The authors acknowledge funding from the European Union’s Horizon 2020 research and innovation program under Marie Sklodowska-Curie Grant Agreement No. 754303 and No. 860060 “Magnetism and the effect of Electric Field” (MagnEFi). J.P.G. also thanks the Laboratoire d\textquotesingle Excellence LANEF in Grenoble (ANR-10-LABX-0051) for its support. B. Fernandez, T. Crozes, Ph. David, E. Mossang and E. Wagner are acknowledged for their technical help.
\end{acknowledgments}
\end{document}